\documentclass[pra,twocolumn,lengthcheck]{revtex4}

\usepackage{amsmath}
\usepackage{amssymb}
\usepackage{amsthm}
\usepackage{graphicx}

\newtheorem{theorem}{Theorem}
\newtheorem{lemma}[theorem]{Lemma}
\newtheorem{conjecture}[theorem]{Conjecture}

\theoremstyle{definition}
\newtheorem{example}[theorem]{Example}
\newtheorem{aside}[theorem]{Aside}

\newcommand{\bra}[1]{\langle{#1}\vert}
\newcommand{\ket}[1]{\vert{#1}\rangle}
\newcommand{\braket}[2]{\langle{#1}\vert{#2}\rangle}
\newcommand{\bi}[1]{\textbf{\textit{#1}}}

\begin{document}

\title{Geometry of Generalized Depolarizing Channels}

\author{Christian K.\ \surname{Burrell}}
\affiliation{Department of Mathematics, Royal Holloway University of London, Egham, Surrey, TW20 0EX, UK}

\begin{abstract}
A generalized depolarizing channel acts on an $N$-dimensional quantum system to compress the ``Bloch ball'' in \mbox{$N^2-1$} directions; it has a corresponding compression vector.  We investigate the geometry of these compression vectors and prove a conjecture of Dixit and Sudarshan \cite{DixSud08}, namely that when \mbox{$N=2^d$} (i.e. the system consists of $d$ qubits) and we work in the Pauli basis then the set of all compression vectors forms a simplex.  We extend this result by investigating the geometry in other bases; in particular we find precisely when the set of all compression vectors forms a simplex.
\end{abstract}

\maketitle

\section{Introduction}\label{sec:intro}

Perhaps the simplest model of noise in a quantum system is that of the isotropic depolarizing channel $\Phi_p$ where with probability probability \mbox{$1-p$} a quantum state $\rho$ is left untouched while with probability $p$ it is mapped to the completely mixed state $\mathbb{I}/N$
\begin{equation*}
 \Phi_p (\rho) = p \frac{\mathbb{I}}{N} + (1-p)\rho
\end{equation*}
This channel results in the Bloch ``ball'' being compressed isotropically by a factor of \mbox{$1-p$}.

One can imagine a slightly more complicated noise model whereby the noise compresses the Bloch ``ball'' anisotropically along axes which are defined by the basis we choose to work in and it is this setting which we investigate below.  The amounts by which we compress along each axis are called the \textbf{compression coefficients} and these form the components of the \textbf{compression vector}.  We investigate the geometric properties of the set of all compression vectors and in particular we ask ``when does this set form a simplex?''

These \textbf{generalized depolarizing channels} (also called \textbf{anisotropic depolarizing channels}) form a broad class of channels which can be realized experimentally (see for example \cite{KarRadKon08}); in the single qubit case they include the bit-flip and phase-flip channels.  It is worth noting that \emph{any} quantum channel (including the use of a spin chain as a quantum channel \cite{Bur06,Bos08}) can be turned into a generalized depolarizing channel as follows:  Alice sends one half of a maximally entangled bipartite state down the channel to Bob; the resulting shared state is a mixed bipartite state which is (excepting very special cases) not maximally entangled; Alice and Bob can now use this shared resource to attempt teleportation \cite{Ben-et-al93} of an unknown $N$-dimensional quantum state $\rho$; the teleportation protocol then acts as a generalized depolarizing channel on this state $\rho$ \cite{BowBos01,VenBosRon07}.

This work is organized as follows: Section~\ref{sec:problem} gives some precise definitions and describes in detail the problem we solve.  Sections~\ref{sec:p-basis}, \ref{sec:gm-basis} and \ref{sec:hw-basis} present the solution to this problem in the Pauli, Gell-Mann and Heisenberg-Weyl bases respectively (we prove the conjecture by Dixit and Sudarshan in section~\ref{sec:p-basis}) whilst section~\ref{sec:other-bases} gives the general solution in an arbitrary basis.  We then discuss changing basis in section \ref{sec:change-basis} before showing how to adopt our method of solution to deal with more general channels in section~\ref{sec:more-channels}. We conclude in section~\ref{sec:conclusions}.

\section{The Problem}\label{sec:problem}

Throughout this work we let \mbox{$\{M_\alpha\}^{N^2-1}_{\alpha=0}$} be a basis for \mbox{$N \times N$} matrices which satisfies the following conditions:
\begin{enumerate}
 \item $M_0=\mathbb{I}$
 \item $\textrm{tr}(M_\alpha)=0$ for all $\alpha \neq 0$
 \item $\textrm{tr}(M^\dag_\alpha M_\beta) = 0$ for all $\alpha \neq \beta$
\end{enumerate}
Abusing terminology slightly (note that \mbox{$\textrm{tr}\,M_0 \neq 0$}), we call such a basis \textbf{trace-free} and \textbf{trace-orthogonal}.  We need not restrict ourselves to trace-orthonormal bases as we can simply divide by $\sqrt{\textrm{tr}(M_\alpha^\dag M_\alpha)}$ when necessary to normalize the basis.

If $\rho$ is the density matrix of any $N$-dimensional quantum system, then we may write
\begin{equation}
 \rho = \frac{1}{N} \left( \mathbb{I} + \sum_{\alpha=1}^{N^2-1} \sqrt{\frac{N(N-1)}{\textrm{tr}(M^\dag_\alpha M_\alpha)}} a_\alpha M_\alpha \right)
\end{equation}
where we have chosen the normalization such that
\begin{itemize}
 \item \mbox{$\| \bi{a} \|  = 1$} if and only if $\rho$ is a pure state
 \item \mbox{$\| \bi{a} \|  < 1$} if and only if $\rho$ is a mixed state
\end{itemize}
where the norm of $\bi{a}$ is defined as \mbox{$\| \bi{a} \| \equiv \sum_{\alpha=1}^{N^2-1} \; |a_\alpha|^2$}.  For notational convenience we add an extra $zero^{th}$ component, \mbox{$a_0 \equiv 1$}, which does not affect the value of $\|\bi{a}\|$.  We call \mbox{$\bi{a}=(a_0,\ldots,a_{N^2-1})$} the \textbf{polarization vector} and $a_\alpha$ the \textbf{polarization coefficients} of the state $\rho$ \textbf{with respect to the basis $\{M_\alpha\}$}.

The \textbf{Bloch ``ball''} is the set of all polarization vectors corresponding to quantum states.  It is important to note that --- except for the single qubit case --- the Bloch ``ball'' is not the ball of unit radius, but rather a convex subset of this ball (this is because for \mbox{$N>2$} some vectors lying within the unit ball are not valid polarization vectors as they do not correspond to positive states).

We can now define a \textbf{(generalized) depolarizing channel with respect to the basis $\{M_\alpha\}$} to be a map $\Phi_\bi{v}$ which satisfies
\begin{enumerate}
 \item $\Phi_\bi{v}$ is a trace-preserving completely positive map
 \item $\Phi_\bi{v}$ compresses the Bloch ``ball'' in the following manner:
  \begin{equation*}
   \Phi_\bi{v}(\rho) = \frac{1}{N} \left( \mathbb{I} + \sum_{\alpha=1}^{N^2-1} \sqrt{\frac{N(N-1)}{\textrm{tr}(M^\dag_\alpha M_\alpha)}} v_\alpha a_\alpha M_\alpha \right)
  \end{equation*}
\end{enumerate}

We call \mbox{$\bi{v}=(v_0,\ldots,v_{N^2-1})$} the \textbf{compression vector} of the channel $\Phi_\bi{v}$ as it specifies the amount by which $\Phi_\bi{v}$ compresses the Bloch ``ball'' along each axis; the $v_\alpha$ are called \textbf{compression coefficients}.  Again, for notational convenience, we have added a $zero^{th}$ component \mbox{$v_0\equiv1$} ($v_0$ is the compression coefficient for \mbox{$M_0 = \mathbb{I}$}, so \mbox{$v_0=1$} ensures that $\Phi_\bi{v}$ is trace-preserving).  

Note that \mbox{$|v_\alpha| \leq 1$} for all $\alpha$.  To see this, let the largest possible magnitude of the polarization coefficient $a_\alpha$ be \mbox{$\widetilde{a}_\alpha=\sup_\rho \{|a_\alpha|\}$} and let $\sigma$ be a state with \mbox{$|a_\alpha| = \widetilde{a}_\alpha$} (so $\sigma$ lies on the boundary of the Bloch ``ball'').  If \mbox{$|v_\alpha|>1$} then $\Phi_\bi{v}(\sigma)$ is not a state (it lies outside the ``Bloch ball''), so \mbox{$|v_\alpha| \leq 1$} as claimed.

It is important to note that the notion of generalized depolarizing channels is highly basis dependent: we \emph{must} define such channels with respect to a given basis.

We make the following observation
\begin{itemize}
 \item If \mbox{$M_\alpha = \gamma M_\beta^\dag$} then hermiticity of $\rho$ tells us that \mbox{$a_\alpha \gamma = a_\beta^*$} (where $a^*$is the complex conjugate of $a$).  Furthermore, since $\Phi_\bi{v}(\rho)$ must be a quantum state (and is therefore hermitian), \mbox{$v_\alpha = v_\beta^*$}.
\end{itemize}
and note its special cases:
\begin{itemize}
 \item If \mbox{$M_\alpha = \gamma M_\alpha^\dag$} then \mbox{$a_\alpha \gamma = a_\alpha^*$} and \mbox{$v_\alpha = v_\alpha^* \in \mathbb{R}$}. (Note that $v_\alpha$ can be negative.)

 \item If $M_\alpha$ is hermitian (\mbox{$\gamma=1$}) then \mbox{$a_\alpha \in \mathbb{R}$} and \mbox{$v_\alpha \in \mathbb{R}$}. 

 \item If $M_\alpha$ is skew-hermitian (\mbox{$\gamma=-1$}) then $a_\alpha$ is pure imaginary and \mbox{$v_\alpha \in \mathbb{R}$}.
\end{itemize}

If we work in a fixed basis it is clear that for each depolarizing channel there is a unique compression vector.  It is natural therefore to ask the question ``which vectors $\bi{v}$ are valid compression vectors corresponding to depolarizing channels $\Phi_\bi{v}$?''  From the definition of a depolarizing channel given above it is clear that the answer to this question is ``a vector $\bi{v}$ is a valid compression vector when the induced map $\Phi_\bi{v}$ is completely positive and \mbox{$v_0=1$} ($\Phi_\bi{v}$ is trace-preserving)''.

It is clear from above that all compression vectors $\bi{v}$ which induce depolarizing channels $\Phi_\bi{v}$ lie within the finite region \mbox{$\mathbb{V} = \{ \bi{v} \textrm{ such that }|v_\alpha| \leq 1$ for all $\alpha\}$}.  However we will see that in general the converse is not true: not all vectors in $\mathbb{V}$ induce depolarizing channels.

To help us decide which induced maps $\Phi_\bi{v}$ are completely positive we employ the \textbf{Choi-Jamiolkowski representation}
\begin{equation*}
 J(\Phi) = \sum_{j,k=0}^{N-1} \Phi(\ket{j}\bra{k}) \otimes \ket{j}\bra{k}
\end{equation*}
where \mbox{$\{\ket{0},\ldots,\ket{N-1}\}$} is the computational basis of our $N$-dimensional quantum system.

We will make use of the following theorem and lemma \begin{theorem}\label{thm:cj}
 A map $\Phi$ is completely positive if and only if $J(\Phi)$ is positive.
\end{theorem}
\begin{proof}
 See \cite{Cho75,Cav99}.
\end{proof}

\begin{lemma}\label{lem:trace-cj}
 If \mbox{$\Phi(\rho) = \textnormal{tr}(A^\dag\rho)B$} \;where $A$ and $B$ are operators (i.e. \mbox{$N \times N$} matrices) then\; \mbox{$J(\Phi)=B\otimes A^*$}.
\end{lemma}
\begin{proof}
 We can write \mbox{$A=\sum_{l,m=0}^{N-1}a_{lm}\ket{l}\bra{m}$}. Then
 \begin{equation*}
  \begin{array}{rcl}
   J(\Phi) & = & \sum_{j,k=0}^{N-1} \textrm{tr}\left(\sum_{l,m=0}^{N-1}a^*_{ml}\ket{l}\braket{m}{j}\bra{k}\right)B \otimes \ket{j}\bra{k}\\

   & = & \sum_{j,k=0}^{N-1} a^*_{jk}B\otimes\ket{j}\bra{k} = B \otimes A^* \hfill \mbox{\qedhere}
  \end{array}
 \end{equation*}
\end{proof}

Since the basis $\{M_\alpha\}$ is trace-orthogonal, we may write
\begin{equation*}
 \rho = \sum_{\alpha=0}^{N^2-1} \frac{\textrm{tr}(M^\dag_\alpha \rho)}{\textrm{tr}(M^\dag_\alpha M_\alpha)}M_\alpha
\end{equation*}
and also
\begin{equation*}
 \Phi_\bi{v}(\rho) = \sum_{\alpha=0}^{N^2-1} \frac{\textrm{tr}(M^\dag_\alpha \rho)}{\textrm{tr}(M^\dag_\alpha M_\alpha)} v_\alpha M_\alpha
\end{equation*}

Since $\textrm{tr}(M^\dag_\alpha M_\alpha)$ is just a constant, a simple application of lemma~\ref{lem:trace-cj} allows us to calculate the Choi-Jamiolkowski representation of the channel $\Phi_\bi{v}$
\begin{equation}\label{eqn:J(Phi)}
 J(\Phi_\bi{v}) = \sum_{\alpha=0}^{N^2-1} v_\alpha \frac{M_\alpha \otimes M^*_\alpha}{\textrm{tr}(M^\dag_\alpha M_\alpha)}
\end{equation}

Using theorem~\ref{thm:cj} we see that $\bi{v}$ is a compression vector corresponding to a completely positive depolarizing channel $\Phi_\bi{v}$ when the eigenvalues of $J(\Phi)$ are all non-negative.  In the following sections we work in different bases and determine which vectors $\bi{v}$ are compression vectors.

\section{Pauli Basis}\label{sec:p-basis}

In this section we restrict ourselves to a multiple qubit setting (\mbox{$N = 2^d$}).  We choose the basis $\{M_\alpha\}$ to be formed of tensor products of the single qubit Pauli matrices: let $\alpha$ be the number whose base-$4$ representation is \mbox{$\alpha_d \alpha_{d-1} \cdots \alpha_1$}
\begin{equation*}
 \alpha = \sum_{j=1}^d 4^{j-1}\alpha_j
\end{equation*}
Then we define the basis matrices $M_\alpha$ by
\begin{equation*}
 M_\alpha = \sigma^{\alpha_1} \otimes \cdots \otimes \sigma^{\alpha_d}
\end{equation*}
where \mbox{$\sigma^0, \sigma^1, \sigma^2, \sigma^3$} are the Pauli spin matrices \mbox{$\mathbb{I}, X, Y, Z$}.  This basis is trace-free and trace-orthogonal. Note that each $M_\alpha$ is hermitian and so all $a_\alpha$ and $v_\alpha$ are real.

In this case it is a simple matter to find the eigenvectors and eigenvalues of $J(\Phi)$.  Since $J(\Phi)$ is an \mbox{$N^2 \times N^2$} matrix (it is a \emph{superoperator}), we can think of it as being an operator on $2d$ qubits (as \mbox{$N^2 = 2^{2d}$}). Let
\begin{equation*}
 \ket{\Psi_{nm}}_j = \sum_{r=0,1} (-1)^{rm} \ket{r}_j \otimes \ket{r + n (\textrm{mod} 2)}_{j+d}
\end{equation*}
(where \mbox{$n,m \in \{0,1\}$}) be Bell states on qubits $j$ and \mbox{$j+d$}. (Note: in more usual notation we have \mbox{$\ket{\Psi_{00}} = \ket{\Phi^+}$}, \mbox{$\ket{\Psi_{01}} = \ket{\Phi^-}$}, \mbox{$\ket{\Psi_{10}} = \ket{\Psi^+}$} and \mbox{$\ket{\Psi_{11}} = \ket{\Psi^-}$}). Observe
\begin{equation*}
 \begin{array}{rcr}
  \sigma^0_j \otimes (\sigma^0_{j+d})^* \ket{\Psi_{nm}}_j & = & \ket{\Psi_{nm}}_j\\

  \sigma^1_j \otimes (\sigma^1_{j+d})^* \ket{\Psi_{nm}}_j & = & (-1)^m\ket{\Psi_{nm}}_j\\

  \sigma^2_j \otimes (\sigma^2_{j+d})^* \ket{\Psi_{nm}}_j & = & (-1)^{n+m}\ket{\Psi_{nm}}_j\\

  \sigma^3_j \otimes (\sigma^3_{j+d})^* \ket{\Psi_{nm}}_j & = & (-1)^n\ket{\Psi_{nm}}_j\\
  \end{array}
\end{equation*}
which may be summarized as
\begin{equation*}
 \sigma^\alpha_j \otimes (\sigma^\alpha_{j+d})^* \ket{\Psi_{nm}}_j = (-1)^{f(\alpha,n,m)} \ket{\Psi_{nm}}_j
\end{equation*}
where
\begin{equation*}
 f(\alpha,n,m) = \left\lfloor \frac{\alpha}{2} \right\rfloor n + \left\lfloor \frac{\alpha+1}{2} \right\rfloor m
\end{equation*}

It is now a straight forward matter to check that the eigenvectors of the Choi-Jamiolkowski representation of an induced channel $\Phi_\bi{v}$ are
\begin{equation*}
 \ket{J_{\bi{nm}}} = \bigotimes_{j=1}^d \ket{\Psi_{n_jm_j}}_j \qquad
 \begin{array}{l}
  \bi{n} = (n_1,\ldots,n_d)\\
  \bi{m} = (m_1,\ldots,m_d)\\
 \end{array}
\end{equation*}
with corresponding eigenvalues
\begin{equation}\label{eqn:lambda-gm}
 \lambda_{\bi{nm}} = \sum_{\alpha=0}^{N^2-1} \frac{v_\alpha}{N} (-1)^{\sum_{j=1}^d f(\alpha_j,n_j,m_j)}
\end{equation}

The important thing to notice here is that each $\lambda_\bi{nm}$ is a linear combination of the compression coefficients $v_\alpha$. Since there are \mbox{$2^{2d} = N^2$} different eigenvectors we have found all the eigenvectors and eigenvalues of $J(\Phi)$.

We have therefore shown which vectors $\bi{v}$ induce depolarizing channels $\Phi_\bi{v}$ --- namely those for which all eigenvalues of the Choi-Jamiolkowski representation of $\Phi_\bi{v}$ are non-negative: \mbox{$\lambda_\bi{nm} \geq 0$} for all $\bi{n}$ and $\bi{m}$.  We are now in a position to prove a conjecture of Dixit and Sudarshan \cite{DixSud08}, which we present as the following theorem.
\begin{theorem}\label{thm:dix-sud}
 When \mbox{$N=2^d$} and we work in the Pauli basis, the set of all compression vectors forms a simplex in compression space.
\end{theorem}
\begin{proof}
 Since each $\lambda_\bi{nm}$ is linear in the compression coefficients $v_\alpha$ then the equation \mbox{$\lambda_\bi{nm}=0$} defines a hyperplane in compression space (which is a real Euclidean vector space of dimension \mbox{$N^2-1$}; it has one dimension for each component of the compression vector --- excepting the $zero^{th}$ component which we suppress as it is identically equal to $1$).

 Since $\Phi_\bi{v}$ is completely positive if and only if all eigenvalues $\lambda_\bi{nm}$ are non-negative, the hyperplanes \mbox{$\lambda_\bi{nm}=0$} must enclose precisely the set of of vectors which induce completely positive depolarizing channels $\Phi_\bi{v}$.  In particular the hyperplanes enclose a finite region of compression space and the shape of this enclosed region must therefore be a simplex.
\end{proof}

We now prove a small lemma before finding the \mbox{\textbf{extremal channels}} of the simplex (which are the depolarizing channels whose compression vectors form the vertices of the simplex).
\begin{lemma}
 The eigenvalues of the Choi-Jamiolkowski representation of $\Phi_\bi{v}$ sum to the system dimension, $N$.
\end{lemma}
\begin{proof}
 First recall that the sum of the eigenvalues of a matrix is simply the trace of that matrix. Then
 \begin{equation*}
  \begin{array}{rcl}
   \sum_{\bi{n},\bi{m}} \lambda_\bi{nm} & = & \textnormal{tr}(J(\Phi_\bi{v}))\\

   & = & \textrm{tr}\left(\sum_{j,k=0}^{N-1} \Phi_\bi{v}(\ket{j}\bra{k}) \otimes \ket{j}\bra{k} \right)\\

   & = & \textrm{tr}\left( \sum_{j=0}^{N-1} \Phi_\bi{v}(\ket{j}\bra{j}) \right)\\

   & = & N \; \textrm{tr} \; \Phi_\bi{v}(\frac{\mathbb{I}}{N}) = N \; \textrm{tr} (\frac{\mathbb{I}}{N}) = N\\
  \end{array}
 \end{equation*}
\end{proof}

\begin{theorem}
 The extremal channels are
 \begin{equation*}
  \Phi^{(\alpha)}(\rho) := M^\dag_\alpha \rho M_\alpha \qquad \alpha \in \{0, \ldots, N^2-1\}
 \end{equation*}
\end{theorem}
\begin{proof}
 First note the identity
 \begin{equation*}
  \sigma^\beta \sigma^\alpha \sigma^\beta = (-1)^{g(\alpha,\beta)} \sigma^\alpha \qquad \alpha,\beta \in \{0,1,2,3\}
 \end{equation*}
 where
 \begin{equation*}
 \begin{array}{rcl}
  g(\alpha,\beta) & \equiv & \left\{
   \begin{array}{lll}
    1 \; (\textrm{mod} 2) & \textrm{if} \; (\alpha,\beta) = & (1,2), (1,3), (2,3),\\
    && (2,1),(3,1),(3,2)\\
    0 \; (\textrm{mod} 2) & \textrm{else} 
   \end{array}\right.\\

   \\

   & \equiv & \alpha\beta(\alpha-\beta)/2 \; (\textrm{mod} 2)
  \end{array}
 \end{equation*}

 We now fix $\beta$ and prove that $\Phi^{(\beta)}$ is an extremal channel.  First note that $\Phi^{(\beta)}$ is completely positive (see for example \cite{nielsen-chuang}) and apply the above identity to see that
 \begin{equation*}
  \begin{array}{rcl}
   \Phi^{(\beta)}(\rho) & = & \sum_{\alpha=0}^{N^2-1} \frac{\textrm{tr}(M^\dag_\beta \rho)}{\textrm{tr}(M^\dag_\alpha M_\alpha)} M^\dag_\beta M_\alpha M_\beta\\
   & = & \sum_{\alpha=0}^{N^2-1} \frac{\textrm{tr}(M^\dag_\beta \rho)}{\textrm{tr}(M^\dag_\alpha M_\alpha)} (-1)^{\sum_{j=1}^d g(\alpha_j,\beta_j)} M_\alpha\\
  \end{array}
 \end{equation*}
 It is clear then that $\Phi^{(\beta)}$ is a depolarizing channel whose compression vector has components
 \begin{equation}\label{eqn:Phi-g}
  v_\alpha = (-1)^{\sum_{j=1}^d g(\alpha_j,\beta_j)}
 \end{equation}
 By combining equations (\ref{eqn:lambda-gm}) and (\ref{eqn:Phi-g}) we see that
 \begin{equation*}
  \lambda_\bi{pq} = \sum_{\alpha=0}^{N^2-1} \frac{1}{N} (-1)^{\sum_{j=1}^d g(\alpha_j,\beta_j) + f(\alpha_j,p_j,q_j)}
 \end{equation*}
 It is clear that if, for fixed $\beta$ and for all $\alpha$,
 \begin{equation}\label{eqn:fg-condition}
  s_{\alpha,\beta,\bi{p},\bi{q}} := \sum_{j=1}^d g(\alpha_j,\beta_j) + f(\alpha_j,p_j,q_j) \equiv 0 \; (\textrm{mod} 2)
 \end{equation}
 then \mbox{$\lambda_\bi{pq} = N$}.  We now show that there exist $\bi{p}$ and $\bi{q}$ such that (\ref{eqn:fg-condition}) holds:

 \begin{itemize}
  \item When \mbox{$\alpha = 4^{r-1}$} (i.e. \mbox{$\alpha_r=1$} and \mbox{$\alpha_{j}=0$} for all \mbox{$j\neq r$}) then
  \begin{equation*}
   s_{\alpha,\beta,\bi{p},\bi{q}} \equiv \frac{\beta_r(1-\beta_r)}{2} + q_r \; (\textrm{mod} 2)
  \end{equation*}
  and so (\ref{eqn:fg-condition}) holds when
  \begin{equation}\label{eqn:q_r}
   q_r \equiv \frac{\beta_r(1-\beta_r)}{2}
  \end{equation}

  \item When \mbox{$\alpha = 3 \times 4^{r-1}$} (i.e. \mbox{$\alpha_r=3$} and \mbox{$\alpha_{j}=0$} for all \mbox{$j\neq r$}) then
  \begin{equation*}
   s_{\alpha,\beta,\bi{p},\bi{q}} \equiv \frac{3\beta_r(3-\beta_r)}{2} + p_r \; (\textrm{mod} 2)
  \end{equation*}
  and so (\ref{eqn:fg-condition}) holds when
  \begin{equation}\label{eqn:p_r}
   p_r \equiv \frac{3\beta_r(3-\beta_r)}{2}
  \end{equation}

  \item When \mbox{$\alpha = 2 \times 4^{r-1}$} (i.e. \mbox{$\alpha_r=2$} and \mbox{$\alpha_{j}=0$} for all \mbox{$j\neq r$}) then
  \begin{equation*}
   s_{\alpha,\beta,\bi{p},\bi{q}} \equiv \frac{2\beta_r(2-\beta_r)}{2} + p_r + q_r \; (\textrm{mod} 2)
  \end{equation*}
  and so (\ref{eqn:fg-condition}) holds when $p_r$ and $q_r$ are picked as in (\ref{eqn:q_r}) and (\ref{eqn:p_r}) above.
 \end{itemize}

 We have now shown that there exist $\bi{p}$ and $\bi{q}$ with \mbox{$\lambda_\bi{pq} = N$}.  Since $\Phi^{(\beta)}$ is completely positive then all the eigenvalues of $J(\Phi^{(\beta)})$ are non-negative; they sum to $N$ and we have found one which is equal to $N$; therefore all other eigenvalues are zero
 \begin{equation*}
  \lambda_\bi{pq}=N \quad \textrm{and} \quad\lambda_\bi{nm}=0 \quad\textrm{ for all } (\bi{n},\bi{m}) \neq (\bi{p},\bi{q})
 \end{equation*}
 Clearly then the compression vector $\bi{v}$ of the map $\Phi^{(\beta)}$ lies on all the hyperplanes \mbox{$\lambda_\bi{nm}=0$} except \mbox{$\lambda_\bi{pq}=0$}, and so $\Phi^{(\beta)}$ must be an extremal channel.

 To finish the proof note that there are $N^2$ vertices of the simplex and there are $N^2$ channels of the form $\Phi^{(\beta)}$ so we have found all the extremal channels.
\end{proof}

It is worth pointing out that any compression vector in the simplex can be written as a convex linear combination of the \textbf{extremal compression vectors} (that is, the compression vectors which form the vertices of the simplex).  This implies that any depolarizing channel can be written as a convex linear combination of the extremal channels
\begin{equation*}
 \Phi_\bi{v} = \sum_{\alpha=0}^{N^2-1} p_\alpha \Phi^{(\alpha)}(\rho) \qquad 0 \leq p_\alpha \leq 1 \; ; \; \sum_{\alpha=0}^{N^2-1} p_\alpha =1
\end{equation*}
Conversely, any channel of this form is a depolarizing channel.  Note the following relationship
\begin{equation*}
 v_\alpha = \sum_{\beta=0}^{N^2-1} p_\beta (-1)^{\sum_{j=1}^d g(\alpha_j,\beta_j)}
\end{equation*}
\begin{example}
 For a single qubit (\mbox{$N=2$}) the compression space has dimension 3 and we see that for a general channel $\Phi_\bi{v}$
 \begin{equation*}
  \begin{array}{rcl}
   v_0 & = & p_0 + p_1 + p_2 + p_3 = 1\\
   v_1 & = & p_0 + p_1 - p_2 - p_3\\
   v_2 & = & p_0 - p_1 + p_2 - p_3\\
   v_3 & = & p_0 - p_1 - p_2 + p_3\\
  \end{array}
 \end{equation*}
 and so we have the following correspondence between extremal channels and compression vectors (we suppress the $zero^{th}$ component of $v$ which is always equal to $1$)
 \begin{equation*}
  \begin{array}{rcccl}
   \Phi^{(0)}(\rho) & = & \mathbb{I} \rho \mathbb{I} & \longleftrightarrow & v=(\phantom{-}1,\phantom{-}1,\phantom{-}1)\\
   \Phi^{(1)}(\rho) & = & X \rho X & \longleftrightarrow & v=(\phantom{-}1,-1,-1)\\
   \Phi^{(2)}(\rho) & = & Y \rho Y & \longleftrightarrow & v=(-1,\phantom{-}1,-1)\\
   \Phi^{(3)}(\rho) & = & Z \rho Z & \longleftrightarrow & v=(-1,-1,\phantom{-}1)\\
  \end{array}
 \end{equation*}
 The geometry of these single qubit depolarizing channels is illustrated in figure~\ref{fig:qubit-tetrahedron}
 \begin{figure}
  \includegraphics[width=0.8\columnwidth]{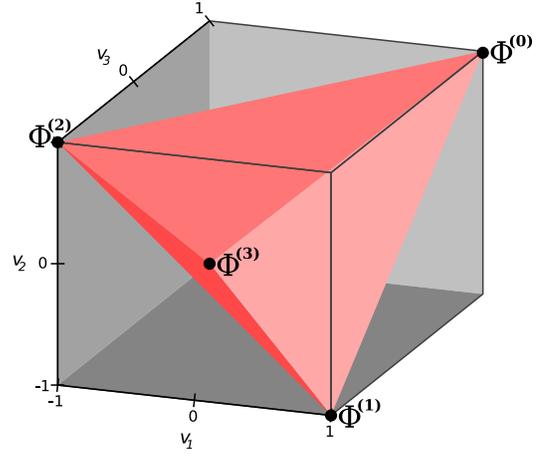}
  \caption{The simplex (a tetrahedron) in compression space corresponding to all single qubit depolarizing channels.}
  \label{fig:qubit-tetrahedron}
 \end{figure}
\end{example}

\section{Gell-Mann Basis}\label{sec:gm-basis}

In the previous section we restricted the dimension of the quantum system to be \mbox{$N=2^d$} so that we could employ the Pauli basis for multiple qubits.  In this section we choose one possible generalization of the Pauli basis, namely the Gell-Mann basis, which allows us to study systems with arbitrary dimension.  The \textbf{Gell-Mann} basis is defined to be
\begin{equation*}
 \begin{array}{rcc}
  X_{jk} &:=& \ket{j}\bra{k} + \ket{k}\bra{j}\\
  Y_{jk} &:=& -i(\ket{j}\bra{k} - \ket{k}\bra{j} )\\
  Z_l &:=& \sqrt{\tfrac{2}{l(l+1)}} \left( \sum_{r=0}^{l-1} \ket{r}\bra{r} - l\ket{l}\bra{l} \right)\\
 \end{array}
\end{equation*}
where \mbox{$j\in\{0,\ldots,N-2\}$}, \; \mbox{$k,l\in\{1,\ldots,N-1\}$} and \mbox{$j<k$}. For notational consistency we identify
\begin{equation*}
 \begin{array}{rclcl}
  M_0 &=& \mathbb{I} && \\
  M_\alpha &=& Z_\alpha && 1 \leq \alpha \leq N-1 \\
  M_\alpha &=& X_{jk} && \alpha = N(1+2j) + 2k - (j+1)(j+2)\\
  M_\alpha &=& Y_{jk} && \alpha = N(1+2j) + 2k - (j+1)(j+2) + 1\\
 \end{array}
\end{equation*}
i.e. \mbox{$\{M_\alpha\} = \{\mathbb{I},Z_1,\ldots,Z_{N-1},X_{01}, Y_{01},\ldots,Y_{N-2,N-1}\}$}.  Note that this basis is trace-free and trace-orthogonal (and reduces to the Pauli basis when \mbox{$N=2$}).  Furthermore each $M_\alpha$ is hermitian and so all $a_\alpha$ and $v_\alpha$ are real.

We now attempt to find the eigenvectors and eigenvalues of $J(\Phi_\bi{v})$ and begin by defining
\begin{equation*}
 \ket{J^\pm_{jk}} = \frac{1}{\sqrt{2}} \left( \ket{j,k} \pm \ket{k,j} \right) 
\end{equation*}
where \mbox{$j,k\in\{0,\ldots,N-1\}$} and \mbox{$j<k$}.  Then it is a simple matter to check that
\begin{equation*}
 \begin{array}{rcl}
  \mathbb{I} \otimes \mathbb{I}^* \ket{J^\pm_{jk}} & = & \ket{J^\pm_{jk}}\\

  &&\\

  Z_l \otimes Z^*_l \ket{J^\pm_{jk}} & = & \left\{\begin{array}{rcl}
   0 \ket{J^\pm_{jk}} & \quad & l<k\\
   \frac{-2}{l+1} \ket{J^\pm_{jk}} && l=k\\
   \frac{2}{l(l+1)} \ket{J^\pm_{jk}} && l>k\\
  \end{array}\right.\\

  &&\\

  X_{lm} \otimes X^*_{lm} \ket{J^\pm_{jk}} & = & \left\{\begin{array}{rcl}
   \pm \ket{J^\pm_{jk}} & \quad & l=j \textrm{ and } m=k\\
   0 \ket{J^\pm_{jk}} && \textrm{else}\\
  \end{array}\right.\\

  &&\\

  Y_{lm} \otimes Y^*_{lm} \ket{J^\pm_{jk}} & = & \left\{\begin{array}{rcl}
   \mp \ket{J^\pm_{jk}} & \quad & l=j \textrm{ and } m=k\\
   0 \ket{J^\pm_{jk}} && \textrm{else}\\
  \end{array}\right.\\
 \end{array}
\end{equation*}
It is clear therefore that $\ket{J^\pm_{jk}}$ are eigenvectors of $J(\Phi_\bi{v})$ and that the corresponding eigenvalues $\lambda^\pm_{jk}$ are linear in the compression coefficients $v_\alpha$.  Now, $J(\Phi_\bi{v})$ has $N^2$ eigenvalues and we have found \mbox{$N^2-N$} of them; let the remaining $N$ eigenvalues be $\lambda_j$ (\mbox{$j\in\{0,\ldots,N-1\}$}).  In order to establish which vectors in compression space induce depolarizing channels we must now find when the remaining $N$ eigenvalues of $J(\Phi_v)$ are non-negative.

Since \mbox{$M_\alpha \otimes M^*_\alpha$} is symmetric for all $\alpha$, $J(\Phi_\bi{v})$ is also symmetric and consequently has real eigenvalues.  By considering the matrix elements of $J(\Phi_\bi{v})$ carefully we see that the only non-zero entries in the two columns indexed by $\bra{j,k}$ and $\bra{k,j}$ (with \mbox{$j<k$}) are in the two rows indexed by $\ket{j,k}$ and $\ket{k,j}$.  We may conjugate $J(\Phi_\bi{v})$ by a permutation matrix $P$ (to permute the rows and columns) to form a matrix $J^\prime(\Phi_v)$ which has identical eigenvalues to $J(\Phi_\bi{v})$.  By repeatedly conjugating by permutation matrices we can block-diagonalize $J(\Phi_\bi{v})$ to obtain a matrix $\widetilde{J}(\Phi_\bi{v})$ which has the following structure
\begin{equation*}
 \widetilde{J}(\Phi_\bi{v}) = \left( \begin{array}{cccc}
  [2 \times 2]     &&& \\
  & \ddots          && \\
  && [2 \times 2]    & \\
  &&& [K(\Phi_\bi{v})] \\
 \end{array} \right)
\end{equation*}
There are \mbox{$N(N-1)/2$} blocks of size \mbox{$2 \times 2$} (each of which has two eigenvalues, $\lambda^+_{jk}$ and $\lambda^-_{jk}$, for some $j$ and $k$) and a large block --- which we call $K(\Phi_\bi{v})$ --- of size \mbox{$N \times N$} (which has eigenvalues \mbox{$\lambda_0, \ldots, \lambda_{N-1}$}).  

By considering the characteristic equation of $K(\Phi_\bi{v})$ we see that
\begin{equation*}
  0 = (t-\lambda_0) \cdots (t-\lambda_{N-1}) = \sum_{j=0}^N (-1)^j t^{N-j} S_j
\end{equation*}
where we have defined the \textbf{eigenvalue sums} $S_j$ to be
\begin{equation*}
 S_0:=1 \quad ; \quad S_j := \sum_{0 \leq k_1 < \cdots < k_j \leq N-1} \lambda_{k_1}\cdots \lambda_{k_j}
\end{equation*}

The following lemma proves that all the eigenvalues \mbox{$\lambda_0,\ldots, \lambda_{N-1}$} are non-negative precisely when all the eigenvalue sums \mbox{$S_0,\ldots, S_N$} are non-negative.

\begin{lemma}
 \mbox{$\lambda_j \geq 0$} for all \mbox{$j\in\{0,\ldots,N-1\}$} if and only if \mbox{$S_i \geq 0$} for all \mbox{$i\in\{0,\ldots,N\}$}.
\end{lemma}
\begin{proof}
 One way is trivial: if \mbox{$\lambda_j \geq 0$} $\forall j$ then \mbox{$S_i \geq 0$} $\forall i$.

 We prove the converse by contradiction: assume that \mbox{$S_k \geq 0$} for all \mbox{$k\in\{1,\ldots,N\}$}, recall that \mbox{$S_0=1$} and note that for any \mbox{$j\in\{0,\ldots,N-1\}$}
 \begin{equation*}
  \begin{array}{rcl}
   S_N & = & \lambda_j (S_{N-1} - \lambda_j(S_{N-2} - \cdots - \lambda_j(S_1-\lambda_j) \cdots))\\

   & = & \sum_{k=1}^N (-1)^{k-1} \lambda_j^k S_{N-k}\\
  \end{array}
 \end{equation*}
 (this is essentially an inclusion-exclusion argument).
 But then we may split this sum up into two terms
 \begin{equation*}
  S_N = \lambda_j \sum_{k\;odd} \lambda_j^{k-1} S_{N-k} - \sum_{k\;even} \lambda_j^k S_{N-k}
 \end{equation*}
  It is clear that these sums over $k$ odd and $k$ even are both non-negative since each $S_j$ is non-negative and $\lambda_j$ appears to an even power in each term.  But then \mbox{$\lambda_j < 0$} implies that \mbox{$S_N < 0$} also, which contradicts the assumption that \mbox{$S_N \geq 0$}, and so \mbox{$\lambda_j \geq 0$}. Since this argument holds for all \mbox{$j\in\{0,\ldots,N-1\}$} we are done.
\end{proof}

Returning to $J(\Phi_\bi{v})$ we see from equation (\ref{eqn:J(Phi)}) that each matrix element consists of a linear combination of the compression coefficients $v_\alpha$ --- a property which is inherited by $K(\Phi_\bi{v})$.  Careful consideration reveals that $S_j$ is a $j^{th}$-order polynomial in the compression coefficients
\begin{equation*}
 S_j = \sum_{0 \leq \alpha_1 \leq \cdots \leq \alpha_j \leq N^2-1} c_{\alpha_1 \cdots \alpha_j} v_{\alpha_1} \cdots v_{\alpha_j}
\end{equation*}
It is tempting to conclude that \mbox{$S_j=0$} is a $j^{th}$-order surface in compression space.  However we must be careful: it is possible that all $j^{th}$-order coefficients $c_{\alpha_1\cdots\alpha_j}$ (i.e. those with \mbox{$\alpha_k \neq 0$} for all $k$) are equal to zero, in which case the degree of the surface \mbox{$S_j=0$} is strictly less than $j$.  For example, when \mbox{$N=2$} the Gell-Mann basis reduces to the Pauli basis and we have already seen in the previous section that all eigenvalues are linear in the compression coefficients; in this case the surface \mbox{$S_2=0$} is a plane and not a quadratic surface.  In general then, we may only conclude that \mbox{$S_j=0$} is a surface of order at most $j$.

We have now proved that the set of all vectors $\bi{v}$ which induce depolarizing channels $\Phi_\bi{v}$ (with respect to the Gell-Mann basis)  form a finite region in compression space which is bounded by
\begin{itemize}
 \item \mbox{$N^2-N+1$} hyperplanes (\mbox{$N^2-N$} are given by \mbox{$\lambda^\pm_{jk}=0$} and the remaining one is given by \mbox{$S_1=0$})

 \item A surface with order at most $2$ (\mbox{$S_2=0$})

 \item[$\vdots$]

 \item A surface with order at most $N$ (\mbox{$S_N=0$})
\end{itemize}

Note that the above does not tell us whether the eigenvalues \mbox{$\lambda_0, \ldots, \lambda_{N-1}$} are linear in the compression coefficients.   It turns out that when \mbox{$N>2$} at least one of the $\lambda_j$ must be non-linear in the compression coefficients --- a fact which is proved in lemma~\ref{lem:gm-curves} in section~\ref{sec:other-bases}.

Whilst it is a slight disappointment that we have been unable to explicitly give expressions for all the eigenvalues of $J(\Phi_\bi{v})$ we can draw some solace from the fact that we have simplified the problem somewhat: in order to see if a vector $\bi{v}$ induces a depolarizing channel we must check to see if all the eigenvalues of the Choi-Jamiolkowski representation $J(\Phi_\bi{v})$ of the induced map $\Phi_\bi{v}$ are non-negative.  Without the above results we would have to use a ``brute-force'' algorithm to calculate the $N^2$ eigenvalues of $J(\Phi_\bi{v})$ directly; however the above results enable us to calculate \mbox{$N^2-N$} of these quickly, leaving only $N$ to be calculated by the brute-force algorithm, which is a substantial speed-up.

\section{Heisenberg-Weyl Basis}\label{sec:hw-basis}

In this section we work in the Heisenberg-Weyl basis which is an alternative generalization of the single-qubit Pauli basis to arbitrary dimension.  It has certain advantages over the Gell-Mann basis (for example, all the Heisenberg-Weyl basis matrices are invertible while most of the Gell-Mann basis matrices are singular).  However there is a price to pay for such convenience which is that the Heisenberg-Weyl basis is not hermitian and so the compression space is a complex vector space.

Let us first define
\begin{equation*}
 X := \sum_{j=0}^{N-1} \ket{j}\bra{j+1} \quad ; \quad
 Z := \sum_{j=0}^{N-1} \omega^j \ket{j}\bra{j}
\end{equation*}
where \mbox{$\omega=e^{2\pi i/N}$} is the primitive $N^{th}$-root of unity and we work modulo $N$.  Note that for \mbox{$r\in\{0,\ldots,N-1\}$}
\begin{equation*}
 X^r = \sum_{j=0}^{N-1} \ket{j}\bra{j+r} \quad ; \quad
 Z^r = \sum_{j=0}^{N-1} \omega^{jr} \ket{j}\bra{j}
\end{equation*}
and that the inverse of $X^r$ is $X^{-r}$ and similarly for $Z^r$.

We now define the \textbf{Heisenberg-Weyl} basis to be
\begin{equation*}
 M_\alpha = M_{jk} := X^j Z^k \quad
 \begin{array}{rl}
  \textrm{where } & j,k\in\{0,\ldots,N-1\}\\
  \textrm{and }   & \alpha = jN+k\\
 \end{array}\hfill
\end{equation*}
Note that this basis is trace-free and trace-orthogonal.

We can now find the eigenvectors and eigenvalues of $J(\Phi_\bi{v})$ by defining
\begin{equation*}
 \ket{J_{jk}} = \sum_{r=0}^{N-1} \omega^{kr} \ket{r,r+j}
\end{equation*}
Then it is a simple matter to check that
\begin{equation*}
 M_{jk} \otimes M_{jk}^* \ket{J_{lm}} = \omega^{mj-kl} \ket{J_{lm}}
\end{equation*}
which shows us that $\ket{J_{lm}}$ are eigenvectors of $J(\Phi_\bi{v})$.  Furthermore, there are $N^2$ distinct $\ket{J_{lm}}$ and we have therefore found all the eigenvectors of $J(\Phi_\bi{v})$.  Relabeling the compression coefficients $v_\alpha$ as $v_{jk}$ (with \mbox{$\alpha = Nj + k$}) we see that the eigenvalues of $J(\Phi_\bi{v})$ are
\begin{equation}\label{eqn:lambda-hw}
 \lambda_{lm} = \sum_{j,k=0}^{N-1} \frac{v_{jk} \omega^{mj-kl}}{N}
\end{equation}

It is easy to check that \mbox{$ X^r Z^s = \omega^{rs} Z^s X^r$} and also \mbox{$M^\dag_{jk} = \omega^{-jk}M_{-j,-k}$}. Therefore, by the observation in the introduction, \mbox{$v_{jk} = v^*_{-j,-k}$}.

In the preceding sections we have had one (real) compression coefficient for each basis matrix. In this section we have seen that most of the compression coefficients come in complex-conjugate pairs corresponding to two basis matrices which are, up to a complex multiplicative factor, hermitian conjugates of each other.  Note that we still have the same number of free compression parameters (since for each pair of conjugate basis matrices there are two free parameters in the associated compression coefficient, namely the real and imaginary parts); and so the compression space (although now complex) still has dimension \mbox{$N^2-1$} (recall that we suppress the dimension corresponding to $v_0$ as it is identically equal to $1$).
\begin{aside}
 As an aside, we find the exact structure of the compression space. We use \mbox{$M_{jk}^\dag=\omega^{-jk}M_{-j,-k}$} to see that \mbox{$M_{jk}=\gamma M_{jk}^\dag$} precisely when \mbox{$j \equiv -j (\textrm{mod }N)$} and \mbox{$k \equiv -k (\textrm{mod }N)$}.

 When $N$ is odd the only solution is \mbox{$j=0$} and \mbox{$k=0$} (i.e. \mbox{$M_{jk}=\mathbb{I}$}).  In this case compression space consists \mbox{$(N^2-1)/2$} complex planes.

 When $N$ is even there are four solutions, namely \mbox{$j,k\in\{0,N/2\}$}.  In this case, compression space consists of $3$ real axes and \mbox{$(N^2-4)/2$} complex planes.
\end{aside}

We now prove that, when working in the Heisenberg-Weyl basis, the set of all compression vectors forms a simplex.  We then find the extremal channels whose compression vectors lie at the vertices of this simplex.

\begin{theorem}
 When we work in the Heisenberg-Weyl basis, the set of all compression vectors forms a simplex in compression space.
\end{theorem}
\begin{proof}
 It is evident from equation (\ref{eqn:lambda-hw}) that the eigenvalues of $J(\Phi_\bi{v})$ are linear in the compression coefficients and so \mbox{$\lambda_{lm}=0$} defines a hyperplane in compression space.

 Since $\Phi_\bi{v}$ is completely positive if and only if all eigenvalues $\lambda_{lm}$ are non-negative, the hyperplanes \mbox{$\lambda_{lm}=0$} must enclose precisely the set of vectors which induce completely positive depolarizing channels $\Phi_\bi{v}$. In particular the hyperplanes enclose a finite region of compression space and the shape of this enclosed region must therefore be a simplex.
\end{proof}

\begin{theorem}
 The extremal channels are
 \begin{equation*}
  \Phi^{(jk)}(\rho) := M^\dag_{jk} \rho M_{jk} \qquad j,k \in \{0, \ldots, N-1\}
 \end{equation*}
\end{theorem}
\begin{proof}
 We fix $j$ and $k$ and prove that $\Phi^{(jk)}$ is extremal.  Using the identity \mbox{$M_{jk}^\dag M_{lm} M_{jk} = \omega^{lk-mj} M_{lm}$} we see
 \begin{equation*}
  \Phi^{(jk)} (\rho) = \sum_{l,m=0}^{N-1} \frac{\textrm{tr}(M_{lm}\rho)}{N} \omega^{lk-mj} M_{lm}
 \end{equation*}
 and so \mbox{$\Phi^{(jk)}(\rho) = \Phi_\bi{v}(\rho)$} is a depolarizing channel with compression coefficients \mbox{$v_{lm} = \omega^{lk-mj}$}.

 We now show that there exists $p$ and $q$ with \mbox{$\lambda_{pq}=N$}.
 \begin{equation*}
  \lambda_{pq} = \sum_{l,m=0}^{N-1} \frac{\omega^{lk-mj} \omega^{lq-mp}}{N}
 \end{equation*}
 So picking \mbox{$p=-j (\textrm{mod } N)$} and \mbox{$q=-k (\textrm{mod } N)$} ensures that \mbox{$\lambda_{pq}=N$}.

 As before we make use of the fact that the eigenvalues of the Choi-Jamiolkowski representation $J(\Phi^{(jk)})$ sum to the system dimension $N$.  Clearly $\Phi^{(jk)}$ is completely positive and so the eigenvalues of $J(\Phi^{(jk)})$ are non-negative and sum to $N$.  Clearly then
 \begin{equation*}
  \lambda_{pq}=N \quad \textrm{and} \quad\lambda_{lm}=0 \quad \textrm{ for all } (l,m) \neq (p,q)
 \end{equation*}
 Therefore the compression vector corresponding to the channel $\Phi^{(jk)}$ lies on all the hyperplanes except one and so it lies at a vertex of the simplex.

 To finish note that there are $N^2$ vertices and $N^2$ distinct extremal channels, so we have found them all.
\end{proof}

As before (when we were working in the Pauli-basis) we note that any compression vector in the simplex can be written as a convex linear combination of the extremal compression vectors; therefore any depolarizing channel can be written as a convex linear combination of the extremal channels and vice-versa.

\section{Other Bases}\label{sec:other-bases}
Having worked in specific bases in the previous sections we now work in an arbitrary trace-free, trace-orthogonal basis $\{M_\alpha\}$.  We have seen that the set of all compression vectors is a simplex precisely when the eigenvalues of $J(\Phi_\bi{v})$ are linear combinations of the compression coefficients $v_\alpha$.  We now find all bases for which this happens.
\begin{theorem}\label{thm:simplex}
 Let $\{M_\alpha\}$ be a trace-free, trace-orthogonal basis.  Then the set of all compression vectors forms a simplex if and only if
 \begin{equation*}
  \left[ M_\alpha \otimes M_\alpha^* , M_\beta \otimes M_\beta^* \right] = 0 \quad \forall \alpha,\beta\in\{0,\ldots,N^2-1\}
 \end{equation*}
\end{theorem}
\begin{proof}
 The set of all compression vectors forms a simplex if and only if all the eigenvalues of $J(\Phi_\bi{v})$ are linear in the compression coefficients (in which case if $\lambda$ is an eigenvalue of $J(\Phi_\bi{v})$ then \mbox{$\lambda=0$} defines a hyperplane --- which forms one of the sides of the simplex).

 Let $\ket{\lambda}$ be an eigenvector of $J(\Phi_\bi{v})$ with eigenvalue $\lambda$.  Then by equation~(\ref{eqn:J(Phi)}) we see that that $\lambda$ is a linear combination of the compression coefficients $v_\alpha$ if and only if $\ket{\lambda}$ is a simultaneous eigenvector of each \mbox{$M_\alpha \otimes M_\alpha^*$}; this occurs if and only if all the \mbox{$M_\alpha \otimes M_\alpha^*$} are simultaneously diagonalizable, which occurs if and only if \mbox{$[ M_\alpha \otimes M_\alpha^* , M_\beta \otimes M_\beta^*]=0$} for all $\alpha$ and $\beta$.
\end{proof}

Note that the condition of theorem~\ref{thm:simplex} is weaker than \mbox{$[ M_\alpha , M_\beta ]=0$} for all $\alpha$ and $\beta$. For example, \mbox{$M_\alpha M_\beta = e^{i\theta_{\alpha\beta}} M_\beta M_\alpha$} satisfies $[ M_\alpha \otimes M_\alpha^* , M_\beta \otimes M_\beta^*]=0$ but not \mbox{$[ M_\alpha , M_\beta ]=0$}.  Furthermore, it is not hard to show that $M_\alpha M_\beta = e^{i\theta_{\alpha\beta}} M_\beta M_\alpha$ is equivalent to \mbox{$[ M_\alpha \otimes M_\alpha^* , M_\beta \otimes M_\beta^*]=0$}.

We can now prove the outstanding result from section~\ref{sec:gm-basis}, which we present as the following lemma.
\begin{lemma}\label{lem:gm-curves}
  When \mbox{$N>2$} and we work in the Gell-Mann basis then the set of all compression vectors has at least one curved side.  Equivalently, at least one of the eigenvalues of $J(\Phi_\bi{v})$ is non-linear in the compression coefficients $v_\alpha$.
\end{lemma}
\begin{proof}
 It is a simple matter to check that
 \begin{equation*}
  \left[ X_{01} \otimes X_{01}^* , X_{02} \otimes X_{02}^* \right] \neq 0
 \end{equation*}
 Then apply theorem~\ref{thm:simplex} above.
\end{proof}

For the remainder of this section we restrict ourselves to working in a basis which is unitary: \mbox{$M_\alpha^{-1}=M_\alpha^\dag$} for all $\alpha$.  When we do this we are able to find the extremal channels whose compression vectors lie at the vertices of the simplex.  Before proving this we first define
\begin{equation*}
 \ket{\Psi} = \sum_{j=0}^{N-1} \ket{j}\ket{j} \quad ; \quad \ket{\lambda_\alpha} = \left( M_\alpha \otimes \mathbb{I} \right) \ket{\Psi}
\end{equation*}
Note that $\ket{\Psi}$ is a maximally entangled state and so the $\ket{\lambda_\alpha}$ are too (as the $M_\alpha$ are unitary).
\begin{lemma}
 If \mbox{$M_\alpha M_\beta = e^{i\theta_{\alpha\beta}} M_\beta M_\alpha$} for all $\alpha$ and $\beta$ and if $M_\alpha$ is unitary for all $\alpha$ then $\ket{\lambda_\alpha}$ are all the eigenstates of $J(\Phi_\bi{v})$.
\end{lemma}
\begin{proof}
 Note that $\ket{\lambda_\alpha} = \sum_{j,k=0}^{N-1} \left( M_\alpha \right)_{j,k} \ket{j}\ket{k}$. Note also that unitarity of $M_\alpha$ ensures that \mbox{$\theta_{\alpha\beta} = - \theta_{\beta\alpha}$}.  Then
 \begin{equation*}
  M_\beta \otimes M_\beta^* \ket{\lambda_\alpha} = \sum_{j,k=0}^{N-1} \left( M_\beta M_\alpha M_\beta^\dag \right) \ket{j}\ket{k} = e^{-i\theta_{\alpha\beta}}\ket{\lambda_\alpha}
 \end{equation*}
 and it is easy to see that 
 \begin{equation*}
  J(\Phi_\bi{v}) \ket{\lambda_\alpha} = \sum_{\beta=0}^{N^2-1} \frac{v_\beta e^{-i\theta_{\alpha\beta}}}{\textrm{tr}(M_\beta^\dag M_\beta)}\ket{\lambda_\alpha} =: \lambda_\alpha \ket{\lambda_\alpha}
 \end{equation*}
 So $\ket{\lambda_\alpha}$ is indeed an eigenstate of $J(\Phi_\bi{v})$.  Finally note that there are $N^2$ eigenstates $\ket{\lambda_\alpha}$ so we have found all the eigenstates of $J(\Phi_\bi{v})$.
\end{proof}

The technical lemma above allows us to find the extremal channels of the simplex:

\begin{theorem}
 If \mbox{$M_\alpha M_\beta = e^{i\theta_{\alpha\beta}} M_\beta M_\alpha$} for all $\alpha$ and $\beta$ and if $M_\alpha$ are all unitary then the extremal channels are
 \begin{equation*}
  \Phi^{(\alpha)}(\rho) \equiv M_\alpha \rho M_\alpha^\dag
 \end{equation*}
\end{theorem}
\begin{proof}
 First note that $\Phi^{(\alpha)}(\rho)$ is completely positive and trace-preserving, so the eigenvalues of $J(\Phi^{(\alpha)})$ are all positive and sum to $N$.  We can expand
 \begin{equation*}
  \Phi^{(\alpha)}(\rho) = \sum_{\beta=0}^{N^2-1} \frac{\textrm{tr}(M_\beta^\dag\rho)}{\textrm{tr}(M_\beta^\dag M_\beta)} e^{i\theta_{\alpha\beta}} M_\beta
 \end{equation*}
 which shows that $\Phi^{(\alpha)}$ is a depolarizing channel with compression coefficients \mbox{$v_\beta = e^{i\theta_{\alpha\beta}}$}.  Now observe that
 \begin{equation*}
  \lambda_\alpha = \sum_{\beta=0}^{N^2-1} \frac{e^{i\theta_{\alpha\beta}}e^{-i\theta_{\alpha\beta}}}{N} = N
 \end{equation*}
 and so \mbox{$\lambda_\beta=0$} for all \mbox{$\beta\neq\alpha$}.  Clearly then $\Phi^{(\alpha)}$ is an extremal channel as its compression vector lies on all the hyperplanes \mbox{$\lambda_\beta = 0$} except \mbox{$\lambda_\alpha =0$}.  To finish, note that there are $N^2$ vertices to the simplex and there are $N^2$ extremal channels $\Phi^{(\alpha)}$ so we have found them all.
\end{proof}

\section{Change of Basis}\label{sec:change-basis}

We now investigate the effect of changing basis, which will enable us to see why in some bases the set of all compression vectors forms a simplex whilst in other bases it does not.

Let \mbox{$\{M_\alpha\}$} and \mbox{$\{L_\alpha\}$} be trace-free, trace-orthogonal bases.  Then we may write
\begin{equation*}
 \frac{M_\alpha}{\sqrt{\textrm{tr}(M_\alpha^\dag M_\alpha)}} = \sum_{\beta=0}^{N^2-1} u_{\alpha\beta} \frac{L_\beta}{\sqrt{\textrm{tr}(L_\beta^\dag L_\beta)}} \quad \textrm{where } u_{\alpha\beta} \in \mathbb{C}
\end{equation*}

Trace-orthogonality of both bases ensures that the \mbox{$N^2 \times N^2$} change-of-basis matrix \mbox{$U=\sum_{\alpha,\beta=0}^{N^2-1} u_{\alpha\beta} \ket{\alpha}\bra{\beta}$} is unitary.  Furthermore if both bases are hermitian then $U$ is actually a real orthogonal matrix.  If we define
\begin{equation*}
 \widehat{\bi{M}} = \left(\frac{M_0}{\sqrt{\textrm{tr}(M_0^\dag M_0)}}, \ldots, \frac{M_{N^2-1}}{\sqrt{\textrm{tr}(M_{N^2-1}^\dag M_{N^2-1})}}\right)
\end{equation*}
and similarly for $\widehat{\bi{L}}$, then we may write \mbox{$\widehat{\bi{M}}=U\widehat{\bi{L}}$}.

We are now in a position to prove
\begin{theorem}\label{thm:qubit-simplices}
 The set of single qubit (\mbox{$N=2$}) compression vectors forms a simplex whenever we work in a hermitian, trace-free and trace-orthogonal basis.

 Furthermore, the extremal channels are conjugations by the basis matrices.
\end{theorem}
\begin{proof}
 We know that any single qubit basis which is hermitian, trace-free and trace-orthogonal can be obtained from the Pauli basis by an orthogonal change of basis.  Let \mbox{$\{L_\alpha\}$} be the Pauli basis and \mbox{$\{M_\alpha\}$} be the new basis.  Then the change of basis matrix has the form
 \begin{equation}
  U = \left(\begin{array}{c|ccc}
             1 & 0 & 0 & 0 \\ \hline
             0 &   &   &   \\
             0 &   & \mathcal{O} & \\
             0 &   &   &   \\
            \end{array}\right)
 \end{equation}
 where $\mathcal{O}$ is a \mbox{$3 \times 3$} orthogonal matrix.  Now, any such matrix $\mathcal{O}$ corresponds to a rotation of the Bloch ball and is equivalent to a unitary conjugation \mbox{$\rho \mapsto V^\dag \rho V$} where $V$ is a \mbox{$2 \times 2$} unitary matrix.  Any such map is completely positive and trace-preserving.  Depolarization with respect to the basis \mbox{$\{M_\alpha\}$} is the same as applying the reverse of the above rotation, followed by depolarization in the Pauli basis, followed by the the above rotation.  Since the set of all compression vectors forms a simplex when we work in the Pauli basis, the set of all compression vectors also forms a simplex when we work in the new basis \mbox{$\{M_\alpha\}$}.

 We know that in the Pauli basis the extremal channels are \mbox{$\Phi^{(\alpha)}(\rho) = L_\alpha^\dag \rho L_\alpha$}.  So when we start in the new basis and rotate to the Pauli basis, the extremal channels will be \mbox{$\Phi^{(\alpha)}(V \rho V^\dag) = L_\alpha^\dag V \rho V^\dag L_\alpha$}; rotating back to the \mbox{$\{M_\alpha\}$} basis (and noting that \mbox{$M_\alpha = V^\dag L_\alpha V$}) we see that the extremal channels are \mbox{$\Phi^{(\alpha)}(\rho) = M_\alpha^\dag \rho M_\alpha$}.
\end{proof}

In the single qubit case it is true that any unitary change of basis $U$ corresponds to an orthogonal rotation of the polarization vector $\bi{a}$; in higher dimensions this is not so ass the following example demonstrates
\begin{example}\label{ex:bloch-rotation}
 Let \mbox{$N=4$} and let
 \begin{equation*}
  \rho = \frac{\mathbb{I}}{4} + \frac{k\sqrt{3}}{4} X \otimes X \qquad k\in \mathbb{R}
 \end{equation*}
 where \mbox{$X=\left(\begin{smallmatrix}0&1\\1&0\end{smallmatrix}\right)$} is the single-qubit Pauli-$X$ matrix.  One can explicitly calculate the eigenvalues of $\rho$; they are \mbox{$\lambda = (1\pm k\sqrt{3})/4$} each with multiplicity two, so $\rho$ is a state when \mbox{$|k| \leq 1/\sqrt{3}$}.

 Let us now apply any change of basis which maps \mbox{$X \otimes X / 2 \mapsto Z_3 / \sqrt{2}$} (that is, we change basis from the hermitian Pauli basis to the hermitian Gell-Mann basis with a unitary change-of-basis matrix $U$).  Then
 \begin{equation*}
  \rho \mapsto \widetilde{\rho} = \frac{\mathbb{I}}{4} + \frac{k\sqrt{6}}{4} Z_3
 \end{equation*}
 which has eigenvalues \mbox{$\lambda = (1+\alpha)/4$} (with multiplicity 3) and \mbox{$\lambda = (1-3\alpha)/4$} (with multiplicity 1).  But then if \mbox{$1/\sqrt{3} \geq k > 1/3$} then $\rho$ is a quantum state but $\widetilde{\rho}$ is not (as it is not positive).
\end{example}

This example demonstrates that some changes of basis do not map the Bloch ``ball'' on to itself.  It is for this reason that the set of all compression vectors forms a simplex in some bases but not in others.

\begin{theorem}
 When two trace-free, trace-orthogonal bases \mbox{$\{M_\alpha\}$} and \mbox{$\{L_\alpha\}$} are related via
 \begin{equation*}
  M_\alpha = V^\dag L_\alpha V
 \end{equation*}
 for all $\alpha$ and some unitary matrix $V$ \emph{(that is, the change-of-basis matrix $U$ maps the Bloch ``ball'' on to itself)}, then the set of compression vectors either forms a simplex in both bases or does not form a simplex in either basis.

 Furthermore, if the extremal channels are conjugations of basis matrices in one basis then they are conjugations of basis matrices in the other basis too.
\end{theorem}
\begin{proof}
 One may adapt the proof of theorem \ref{thm:qubit-simplices}.  (Note that the condition \mbox{$M_\alpha = V^\dag L_\alpha V$} for all $\alpha$ guarantees that the ``Bloch ball'' is rotated onto itself and we avoid the problems exhibited in example \ref{ex:bloch-rotation}.)
\end{proof}

\section{More General Channels}\label{sec:more-channels}

Throughout the previous sections of this work we have studied depolarizing channels.  Our method can be extended to deal with a more general class of channels, namely those which both compress the Bloch ``ball'' and then translate it.  Let $\Phi_\bi{v,t}$ be such a channel defined by its action on the polarization vector
\begin{equation*}
 \Phi_\bi{v,t} : \left(
 \begin{array}{c}
  a_0 \\ \vdots \\ a_{N^2-1}
 \end{array}\right) \mapsto \left(
 \begin{array}{c}
  v_0 a_0 \\ \vdots \\ v_{N^2-1} a_{N^2-1}
 \end{array}\right) + \left(
 \begin{array}{c}
  t_0 \\ \vdots \\ t_{N^2-1}
 \end{array}\right)
\end{equation*}
We call the vector \mbox{$\bi{t} = (t_0, \ldots, t_{N^2-1})$} the \textbf{translation vector} and the $t_\alpha$ the \textbf{translation coefficients}.  Note that \mbox{$t_0=0$} to ensure that $\Phi_\bi{v,t}$ is trace-preserving.  So far we have only studied channels \mbox{$\Phi_\bi{v} = \Phi_\bi{v,0}$}, but we now extend to general $\bi{t}$.  We may expand
\begin{equation*}
 \Phi_\bi{v,t} (\rho) = \frac{1}{N} \left( \mathbb{I} + \sum_{\alpha=1}^{N^2-1} \sqrt{\frac{N(N-1)}{\textrm{tr}(M^\dag_\alpha M_\alpha)}} (v_\alpha a_\alpha + t_\alpha) M_\alpha \right)
\end{equation*}

If we now define
\begin{equation*}
 V_\alpha(\rho) := \frac{\textrm{tr}(M_\alpha^\dag \rho)}{\textrm{tr}(M_\alpha^\dag M_\alpha)} M_\alpha
\end{equation*}
for all $\alpha$; and 
\begin{equation*}
 T_\alpha(\rho) := \frac{\textrm{tr}(\mathbb{I}\rho)M_\alpha}{N} \sqrt{\frac{N(N-1)}{\textrm{tr}(M_\alpha^\dag M_\alpha)}}  = \frac{1}{N} \sqrt{\frac{N(N-1)}{\textrm{tr}(M_\alpha^\dag M_\alpha)}} M_\alpha
\end{equation*}
for \mbox{$\alpha\in\{1,\ldots,N^2-1\}$} and \mbox{$T_0(\rho):= 0$} then
\begin{equation*}
 \Phi_\bi{v,t}(\rho) = \sum_{\alpha=0}^{N^2-1} v_\alpha V_\alpha(\rho) + t_\alpha T_\alpha(\rho)
\end{equation*}
Recalling the fact that if a channel $\chi$ has the expression \mbox{$\chi(\rho) = \textrm{tr}(X^\dag \rho)Y$} then the Choi-Jamiolkowski representation is \mbox{$J(\chi) = Y \otimes X^*$} we see that
\begin{equation*}
 J(V_\alpha) = \frac{M_\alpha \otimes M_\alpha^*}{\textrm{tr}(M_\alpha^\dag M_\alpha)} \qquad J(T_\alpha) = \frac{1}{N}\sqrt{\frac{N(N-1)}{\textrm{tr}(M_\alpha^\dag M_\alpha)}} M_\alpha \otimes \mathbb{I}
\end{equation*}
which enables us to find the Choi-Jamiolkowski representation of the channel $\Phi_\bi{v,t}$
\begin{equation*}
 J(\Phi_\bi{v,t}) = \sum_{\alpha=0}^{N^2-1} M_\alpha \otimes \left( \frac{v_\alpha M_\alpha^*}{\textrm{tr}(M_\alpha^\dag M_\alpha)} + \frac{t_\alpha \mathbb{I}}{N}\sqrt{\frac{N(N-1)}{\textrm{tr}(M_\alpha^\dag M_\alpha)}} \right)
\end{equation*}

In order to find which parameters $(\bi{v},\bi{t})$ induce completely positive channels one has to find for which parameters all the eigenvalues of $J(\Phi_\bi{v,t})$ are non-negative. (Recall that these channels are automatically trace-preserving if \mbox{$v_0=1$} and \mbox{$t_0=0$}).

\begin{example}
 Let us work in the Pauli basis and consider a single-qubit channel $\Phi_\bi{v,t}$. We fix the translation vector \mbox{$\bi{t} = (0,0,0,t_z)$} and find the set of all compression vectors $\bi{v}$. Now,
 \begin{equation*}
 \begin{array}{rcl}
  J(\Phi_\bi{v,t}) & = & (\mathbb{I} \otimes \mathbb{I})/2 + v_x (X \otimes X)/2 + v_y (Y\otimes Y)/2 \\
  && + v_z (Z \otimes Z)/2 + t_z (Z \otimes \mathbb{I})/2
 \end{array}
 \end{equation*}
 It turns out that --- for non-zero $t_z$ --- the eigenvectors of $J(\Phi_\bi{v,t})$ are no longer Bell-states, but rather linear combinations of them.  Let
 \begin{equation*}
  \begin{array}{c}
   \ket{\lambda_{0cd}} = c\ket{\Psi_{00}} + d\ket{\Psi_{01}}\\
   \ket{\lambda_{1cd}} = c\ket{\Psi_{10}} + d\ket{\Psi_{11}}
 \end{array}
 \end{equation*}
 where \mbox{$c,d\in\mathbb{R}$} are real numbers, \mbox{$c^2 + d^2 =1$} and $\ket{\Psi_{mn}}$ (\mbox{$m,n\in\{0,1\}$}) are the Bell-states defined in section \ref{sec:p-basis}.  One can easily check that
 \begin{equation*}
  \begin{array}{c}
   J(\Phi_\bi{v,t}) \ket{\lambda_{0cd}} = \lambda_0 \ket{\lambda_{0cd}}\\
   J(\Phi_\bi{v,t}) \ket{\lambda_{1cd}} = \lambda_1 \ket{\lambda_{1cd}}
  \end{array}
 \end{equation*}
 where the eigenvalues satisfy the following relations
 \begin{equation*}
  \begin{array}{c}
   \lambda_0 c = \frac{c}{2}(1+v_x+v_y+v_z) + \frac{d}{2}t_z \\
   \lambda_0 d = \frac{d}{2}(1-v_x-v_y+v_z) + \frac{c}{2}t_z \\
   \lambda_1 c = \frac{c}{2}(1+v_x-v_y-v_z) + \frac{d}{2}t_z \\
   \lambda_1 d = \frac{d}{2}(1-v_x+v_y-v_z) + \frac{c}{2}t_z
  \end{array}
 \end{equation*}
 which can be solved to give
 \begin{equation*}
  \begin{array}{c}
   \lambda^\pm_0 = \frac{1}{2}(1+v_z) \pm \sqrt{(v_x+v_y)^2+t_z^2}\\
   \lambda^\pm_1 = \frac{1}{2}(1-v_z) \pm \sqrt{(v_x-v_y)^2+t_z^2}
  \end{array}
 \end{equation*}
 Note that there are two possible values for $\lambda_0$ and $\lambda_1$: one for \mbox{$d=\sqrt{1-c^2}$} and one for \mbox{$d=-\sqrt{1-c^2}$};  we have therefore found all four eigenvalues of $J(\Phi_\bi{v,t})$.  Clearly for \mbox{$t_z\neq 0$} the surfaces \mbox{$\lambda^\pm_0=0$} and \mbox{$\lambda^\pm_1=0$} are curved. This reproduces the results of \cite{RusSzaWer01}, which also shows that the set of all compression vectors forms an ``asymmetrically rounded tetrahedron''.

 We could have worked through the above example with $t_x$ or $t_y$ non-zero but then the generic form of the eigenvectors of $J(\Phi_\bi{v,t})$ would be more complicated: \mbox{$\ket{\lambda} = c\ket{\Psi_{00}} + d\ket{\Psi_{01}} + e\ket{\Psi_{10}} + f\ket{\Psi_{11}}$} with the coefficients \mbox{$c,d,e,f\in\mathbb{R}$} satisfying \mbox{$c^2+d^2+e^2+f^2=1$}.
\end{example}

Based on the above example we note that even when we work in a basis where the set of all compression vectors forms a simplex for \mbox{$\bi{t} = \textbf{0}$}, a tiny perturbation to \mbox{$\bi{t} \neq \textbf{0}$} destroys this simplex: the set of compression vectors $\bi{v}$ forms a set with a curved boundary; in this situation there are infinitely many extremal channels.

\section{Conclusions}\label{sec:conclusions}

In summary, we defined \emph{trace-free, trace-orthogonal bases} for quantum states and used these to define \emph{depolarizing channels}.  Each depolarizing channel has an associated \emph{compression vector} which lies in \emph{compression space} --- which is an \mbox{$N^2-1$} dimensional vector space (we suppressed the other dimension as the first component of the compression vector is always equal to unity).

We showed that the set of all compression vectors forms a simplex when we work in either the \emph{Pauli basis} or the \emph{Heisenberg-Weyl basis}, but that it does not form a simplex when we work in the \emph{Gell-Mann} basis.  Furthermore, in the Pauli and Heisenberg-Weyl bases we found the extremal channels whose compression vectors lie at the vertices of the corresponding simplex.

Working in a general trace-free, trace-orthogonal basis, we showed precisely for which bases the set of compression vectors forms a simplex.  When the basis matrices were all unitary we were able to find the extremal channels.  We discussed the effects of changing basis and this indicated why the set of compression vectors forms a simplex in some bases but not in others.  Finally we generalized our methods to deal with a class of more general quantum channels.

I would now like to recall a theorem proved in \cite{Wat08} but first I must define a \textbf{doubly stochastic} quantum channel acting on an $N$-dimensional quantum system to be any completely-positive, trace-preserving channel which leaves the completely mixed state untouched: \mbox{$\Phi(\mathbb{I}/N) = \mathbb{I}/N$}.

Let us now define $\mathcal{S}_N$ to be the set of all doubly stochastic channels which act on an $N$-dimensional quantum system.  As $\mathcal{S}_N$ is a convex set we may define $\mathcal{T}_N$ to be the set of extremal channels in $\mathcal{S}_N$.

A quantum channel $\Phi$ is said to be \textbf{mixed unitary} if there exist unitary matrices \mbox{$U_1,\ldots,U_k$} and a probability distribution \mbox{$p_1,\ldots,p_k$} such that $\Phi$ is a convex sum of the unitary conjugation channels
\begin{equation*}
 \Phi(\rho) = \sum_{j=1}^k p_j U_j^\dag \rho U_j
\end{equation*}

It is known that all doubly stochastic single-qubit channels are mixed unitary, but that in higher dimensions (\mbox{$N>2$}) there are doubly stochastic channels which are not mixed unitary.  (They can however be expressed as an affine sum of unitary channels \mbox{$\sum_{j=1}^k \lambda_j U_j^\dag \rho U_j$} where \mbox{$\lambda_1,\ldots,\lambda_k$} are affine parameters: \mbox{$\lambda_j\in\mathbb{R}$} and \mbox{$\sum_j \lambda_j =1$}.)

In particular, when \mbox{$N>2$} and we work in any unitary basis for which the set of compression vectors forms a simplex, there exist doubly stochastic channels which do not lie within the simplex and are therefore not depolarizing channels.  However, the following theorem (proved in \cite{Wat08}) tells us that all such channels when appropriately averaged with the isotropic completely depolarizing channel $\Omega:=\Phi_1$, become mixed unitary.

\begin{theorem}
 Let $\chi$ be any doubly-stochastic quantum channel acting on $N$-dimensional quantum systems.  Then for any \mbox{$0\leq p \leq 1/(N^2-1)$} the channel
 \begin{equation*}
  p\chi + (1-p)\Omega
 \end{equation*}
is mixed unitary.
\end{theorem}

It is tempting to think that $\Omega$ is not the only channel with which one can average to obtain a mixed unitary:
\begin{conjecture}
 Let $\chi$ be any doubly-stochastic quantum channel acting on $N$-dimensional quantum systems and let $\Phi$ be any depolarizing channel whose compression vector lies within a simplex whose vertices are the compression vectors of unitary conjugation depolarizing channels.  Then there exists a constant \mbox{$\delta(\Phi)>0$} such that for \mbox{$0\leq p \leq \delta$} the following channel is mixed unitary
 \begin{equation*}
  p\chi + (1-p)\Phi
 \end{equation*}
\end{conjecture}

\emph{Acknowledgments ---} Helpful conversations with Tobias Osborne are gratefully acknowledged.  This work was funded by EPSRC.


\begin{thebibliography}{12}
\expandafter\ifx\csname natexlab\endcsname\relax\def\natexlab#1{#1}\fi
\expandafter\ifx\csname bibnamefont\endcsname\relax
  \def\bibnamefont#1{#1}\fi
\expandafter\ifx\csname bibfnamefont\endcsname\relax
  \def\bibfnamefont#1{#1}\fi
\expandafter\ifx\csname citenamefont\endcsname\relax
  \def\citenamefont#1{#1}\fi
\expandafter\ifx\csname url\endcsname\relax
  \def\url#1{\texttt{#1}}\fi
\expandafter\ifx\csname urlprefix\endcsname\relax\def\urlprefix{URL }\fi
\providecommand{\bibinfo}[2]{#2}
\providecommand{\eprint}[2][]{\url{#2}}

\bibitem[{\citenamefont{Dixit and Sudarshan}(2008)}]{DixSud08}
\bibinfo{author}{\bibfnamefont{K.}~\bibnamefont{Dixit}} \bibnamefont{and}
  \bibinfo{author}{\bibfnamefont{E.~C.~G.} \bibnamefont{Sudarshan}},
  \bibinfo{journal}{Phys. Rev. A} \textbf{\bibinfo{volume}{78}},
  \bibinfo{pages}{032308} (\bibinfo{year}{2008}).

\bibitem[{\citenamefont{Karpinski et~al.}(2008)\citenamefont{Karpinski,
  Radzewicz, and Banaszek}}]{KarRadKon08}
\bibinfo{author}{\bibfnamefont{M.}~\bibnamefont{Karpinski}},
  \bibinfo{author}{\bibfnamefont{C.}~\bibnamefont{Radzewicz}},
  \bibnamefont{and} \bibinfo{author}{\bibfnamefont{K.}~\bibnamefont{Banaszek}},
  \bibinfo{journal}{J. Opt. Soc. Am. B} \textbf{\bibinfo{volume}{25}},
  \bibinfo{pages}{668} (\bibinfo{year}{2008}).

\bibitem[{\citenamefont{Burgarth}(2006)}]{Bur06}
\bibinfo{author}{\bibfnamefont{D.~K.} \bibnamefont{Burgarth}}, Ph.D. thesis,
  \bibinfo{school}{UCL} (\bibinfo{year}{2006}), \eprint{arXiv: 0704.1309}.

\bibitem[{\citenamefont{Bose}(2007)}]{Bos08}
\bibinfo{author}{\bibfnamefont{S.}~\bibnamefont{Bose}},
  \bibinfo{journal}{Contemporary Physics} \textbf{\bibinfo{volume}{48}},
  \bibinfo{pages}{13} (\bibinfo{year}{2007}).

\bibitem[{\citenamefont{Bennett et~al.}(1993)\citenamefont{Bennett, Brassard,
  Josza, Peres, and Wooters}}]{Ben-et-al93}
\bibinfo{author}{\bibfnamefont{C.~H.} \bibnamefont{Bennett}},
  \bibinfo{author}{\bibfnamefont{G.}~\bibnamefont{Brassard}},
  \bibinfo{author}{\bibfnamefont{C.~C.~R.} \bibnamefont{Josza}},
  \bibinfo{author}{\bibfnamefont{A.}~\bibnamefont{Peres}}, \bibnamefont{and}
  \bibinfo{author}{\bibfnamefont{W.~K.} \bibnamefont{Wooters}},
  \bibinfo{journal}{Phys. Rev. Lett.} \textbf{\bibinfo{volume}{70}},
  \bibinfo{pages}{1895} (\bibinfo{year}{1993}).

\bibitem[{\citenamefont{Bowen and Bose}(2001)}]{BowBos01}
\bibinfo{author}{\bibfnamefont{G.}~\bibnamefont{Bowen}} \bibnamefont{and}
  \bibinfo{author}{\bibfnamefont{S.}~\bibnamefont{Bose}},
  \bibinfo{journal}{Phys. Rev. Lett.} \textbf{\bibinfo{volume}{87}}
  (\bibinfo{year}{2001}).

\bibitem[{\citenamefont{Venuti et~al.}(2007)\citenamefont{Venuti, Boschi, and
  Roncaglia}}]{VenBosRon07}
\bibinfo{author}{\bibfnamefont{L.~C.} \bibnamefont{Venuti}},
  \bibinfo{author}{\bibfnamefont{C.~D.~E.} \bibnamefont{Boschi}},
  \bibnamefont{and}
  \bibinfo{author}{\bibfnamefont{M.}~\bibnamefont{Roncaglia}},
  \bibinfo{journal}{Phys. Rev. Lett.} \textbf{\bibinfo{volume}{99}},
  \bibinfo{pages\bibitem[{\citenamefont{Bennett et~al.}(1993)\citenamefont{Bennett, Brassard,
  Josza, Peres, and Wooters}}]{Ben-et-al93}
\bibinfo{author}{\bibfnamefont{C.~H.} \bibnamefont{Bennett}},
  \bibinfo{author}{\bibfnamefont{G.}~\bibnamefont{Brassard}},
  \bibinfo{author}{\bibfnamefont{C.~C.~R.} \bibnamefont{Josza}},
  \bibinfo{author}{\bibfnamefont{A.}~\bibnamefont{Peres}}, \bibnamefont{and}
  \bibinfo{author}{\bibfnamefont{W.~K.} \bibnamefont{Wooters}},
  \bibinfo{journal}{Phys. Rev. Lett.} \textbf{\bibinfo{volume}{70}},
  \bibinfo{pages}{1895} (\bibinfo{year}{1993}).}{060401} (\bibinfo{year}{2007}).

\bibitem[{\citenamefont{Choi}(1975)}]{Cho75}
\bibinfo{author}{\bibfnamefont{M.}~\bibnamefont{Choi}},
  \bibinfo{journal}{Linear Alg. Appl.} \textbf{\bibinfo{volume}{10}},
  \bibinfo{pages}{285} (\bibinfo{year}{1975}).

\bibitem[{\citenamefont{Caves}(1999)}]{Cav99}
\bibinfo{author}{\bibfnamefont{C.~M.} \bibnamefont{Caves}},
  \bibinfo{journal}{Journal of Superconductivity}
  \textbf{\bibinfo{volume}{12}}, \bibinfo{pages}{707} (\bibinfo{year}{1999}).

\bibitem[{nie(2000)}]{nielsen-chuang}
\emph{\bibinfo{title}{Quantum Computation and Quantum Information}}
  (\bibinfo{publisher}{Cambridge University Press}, \bibinfo{year}{2000}).

\bibitem[{\citenamefont{Ruskai et~al.}(2002)\citenamefont{Ruskai, Szarek, and
  Werner}}]{RusSzaWer01}
\bibinfo{author}{\bibfnamefont{M.~B.} \bibnamefont{Ruskai}},
  \bibinfo{author}{\bibfnamefont{S.}~\bibnamefont{Szarek}}, \bibnamefont{and}
  \bibinfo{author}{\bibfnamefont{E.}~\bibnamefont{Werner}},
  \bibinfo{journal}{Lin. Alg. Appl.} \textbf{\bibinfo{volume}{347}},
  \bibinfo{pages}{159} (\bibinfo{year}{2002}).

\bibitem[{\citenamefont{Watrous}(2009)}]{Wat08}
\bibinfo{author}{\bibfnamefont{J.}~\bibnamefont{Watrous}},
  \bibinfo{journal}{Quantum Information and Computation}
  \textbf{\bibinfo{volume}{9}}, \bibinfo{pages}{406} (\bibinfo{year}{2009}).

\end{thebibliography}
\end{document}